\documentclass[amsmath,amssymb,twocolumn,preprintnumbers]{revtex4}
\input{colordvi.tex}
\usepackage{bm}% bold math
\usepackage[mathscr]{eucal}% Euler script
\usepackage[dvips]{graphicx,color}

\begin{document}

\preprint{YITP-08-89, APCTP Pre2008 - 009}

\title{Non-Gaussianity in the Cosmic Microwave Background
Temperature Fluctuations from Cosmic (Super-)Strings}

\author{
Keitaro Takahashi$^1$, Atsushi Naruko$^2$, Yuuiti Sendouda$^2$,
Daisuke Yamauchi$^2$, Chul-Moon Yoo$^3$ and Misao Sasaki$^2$
}
\affiliation{${}^1$
Department of Physics and Astrophysics, Nagoya University,
Nagoya 464-8602, Japan
}
\affiliation{${}^2$
Yukawa Institute for Theoretical Physics, Kyoto University,
Kyoto 606-8502, Japan
}
\affiliation{${}^3$
Asia Pacific Center for Theoretical Physics,
Pohang University of Science and Technology, Pohang 790-784, Korea
}

\date{\today}

\begin{abstract}
We compute analytically the small-scale temperature fluctuations
of the cosmic microwave background from cosmic (super-)strings
and study the dependence on the string intercommuting probability $P$.
We develop an analytical model which describes the evolution
of a string network and calculate the numbers of
string segments and kinks in a horizon volume.
Then we derive the probability distribution function (pdf)
which takes account of finite angular resolution of observation. 
The resultant pdf consists of a Gaussian part due to frequent scatterings
by long string segments and a non-Gaussian tail due to close
encounters with kinks. The dispersion of the Gaussian part is
reasonably consistent with that obtained by numerical simulations
by Fraisse et al.. On the other hand, the non-Gaussian tail contains
two phenomenological parameters which are determined by comparison
with the numerical results for $P=1$. Extrapolating the pdf
to the cases with $P<1$, we predict that the non-Gaussian
feature is suppressed for small $P$.
\end{abstract}

\maketitle

%%%%%%%%%%%%%%%%%%%%%%%%%%%%%%%%%%%%%%%%%%%%%%%%%%%%%%%%%%%%%%%%%
\section{Introduction}
%%%%%%%%%%%%%%%%%%%%%%%%%%%%%%%%%%%%%%%%%%%%%%%%%%%%%%%%%%%%%%%%%

The imprint of cosmic strings on the cosmic microwave background
(CMB) has been widely studied. Although cosmic strings are
excluded as a dominant source of the observed large-angular-
scale anisotropies \cite{Perivolaropoulos05},
they could still be observable at small scales
\cite{Hindmarsh94,Fraisse07,Pogosian08} with new arcminute
CMB experiments, such as the South Pole Telescope \cite{SPT}
or the Atacama Cosmology Telescope \cite{ACT}.
Because the structure of a string network is highly nonlinear,
it would naturally induce a non-Gaussian feature in the CMB
fluctuations. In particular, a moving straight string produces
discontinuities in CMB temperature, called the Kaiser-Stebbins
effect \cite{KaiserStebbins}, and a temperature gradient map
has been suggested as a means for detecting such an effect
\cite{Gott} (see also \cite{Hammond08}).
The non-Gaussian feature would also appear in the bispectrum
\cite{Gangui}, which has attracted much attention in
the CMB community \cite{Bartolo}. Fraisse et al. \cite{Fraisse07}
found that the probability distribution function (pdf) of
the temperature fluctuations has a non-Gaussian tail and
negative skewness. These non-Gaussian features may help us
distinguish cosmic string signals from other secondary effects
and hence enhance their observability.

Recently, cosmic superstrings have attracted much attention
in the context of inflation in string theory
\cite{Majumdar05,Sakellariadou08}. Cosmic superstrings have
properties different from conventional field-theoretic
cosmic strings. One of the observationally interesting
differences is concerning the intercommuting probability $P$.
It can be significantly smaller than unity for superstrings
while $P = 1$ is normally assumed for field-theoretic strings
(but see \cite{Salmi08}). This difference may be used to
distinguish superstrings from field-theoretic strings
observationally.

In this paper, we compute analytically the pdf of the
small-scale CMB temperature fluctuations and study its
dependence on $P$. At small scales where the primary
fluctuations are damped, only the integrated Sachs-Wolfe
(ISW) effect is relevant. Because the contribution from
loops was shown to be insignificant \cite{Fraisse07},
we focus on the ISW effect of long string segments and kinks.
We first present the basic formulae for the temperature
fluctuations induced by long string segments and kinks
\cite{Stebbins} (section \ref{fluctuation}), and follow
the evolution of the number densities of segments and kinks by
combining and extending a velocity-dependent one-scale model
\cite{Martins,Avgoustidis06} and a kink model \cite{Allen}
(section \ref{scaling}). Then, in section \ref{pdf_CMB},
we derive the pdf showing that the results in
\cite{Fraisse07} can be interpreted with our simple model.
Also the $P$ dependence of the pdf is presented and
the non-Gaussianity is predicted to be suppressed for small $P$.
Finally we summarize our results in section \ref{summary}.

%%%%%%%%%%%%%%%%%%%%%%%%%%%%%%%%%%%%%%%%%%%%%%%%%%%%%%%%%%%%%%%%%
\section{Temperature fluctuations due to cosmic strings
\label{fluctuation}}
%%%%%%%%%%%%%%%%%%%%%%%%%%%%%%%%%%%%%%%%%%%%%%%%%%%%%%%%%%%%%%%%%

First we summarize the basic formulae for the CMB temperature
fluctuations due to cosmic strings, following \cite{Stebbins}.
We denote the position of a cosmic string by
$\vec{r}(t,\sigma)$ where $t$ and $\sigma$ are the time and
position on the string worldsheet, respectively.
The equations of motion and constraints in a flat spacetime
are given by,
\begin{eqnarray}
&& \ddot{\vec{r}} - \vec{r}^{~\prime\prime} = 0, \\
&& |\dot{\vec{r}}|^2 + |\vec{r}^{~\prime}|^2 = 1, \\
&& \dot{\vec{r}} \cdot \vec{r}^{~\prime} = 0,
\end{eqnarray}
where the dot and prime denote the derivatives with respect
to $t$ and $\sigma$, respectively. Photons obtain or lose
their energies due to the gravitational field of cosmic
strings. The temperature fluctuation in the direction
$\hat{n}$, $\Delta \equiv \Delta T/T$, due to a straight
segment with the length $\xi$ is written as
\begin{equation}
\Delta_{\rm seg}(\hat{n})
= 8 \frac{v}{\sqrt{1-v^2}} \alpha_{\rm seg} G \mu
  \arctan{\frac{\xi}{\delta}},
\label{fluctuation_segment_atan}
\end{equation}
where $G$ is the Newton constant, $\mu$ is the string tension,
$\delta$ is the impact parameter of a photon ray,
$v \equiv |\dot{\vec{r}}|$ is the velocity of the segment and
\begin{equation}
\alpha_{\rm seg}
 = \hat{n} \cdot
   \left( \frac{\vec{r}^{~\prime}}{|\vec{r}^{~\prime}|} \times
          \frac{\dot{\vec{r}}}{|\dot{\vec{r}}|}
   \right)
\label{alpha_seg}
\end{equation}
is a factor which represents the configuration of the segment
and the direction of the line of sight and can be positive
or negative. In the limit that the impact parameter is much
smaller than the segment length, $\delta \ll \xi$,
Eq. (\ref{fluctuation_segment_atan}) is reduced to
the well known formula,
\begin{equation}
\Delta_{\rm seg}(\hat{n})
= 4 \pi \frac{v}{\sqrt{1-v^2}} \alpha_{\rm seg} G \mu.
\label{fluctuation_segment}
\end{equation}
In fact, Eq. (\ref{fluctuation_segment_atan}) can be well
approximated by Eq. (\ref{fluctuation_segment}) for
$\delta \lesssim \xi$ while it approaches zero for
$\delta \gtrsim \xi$. Therefore a segment with
the length $\xi$ has an effective cross section $\sim \xi^2$.

On the other hand, a kink can be modeled as a nonsmooth
junction of two straight segments with different directions,
$\vec{r}^{~\prime}$ \cite{Stebbins}. Then the temperature
fluctuation with the impact parameter $\delta$ is
\begin{equation}
\Delta_{\rm kink}(\hat{n})
= - 4 G \mu \alpha_{\rm kink}
    \log{\frac{\delta}{L_{\rm kink}}}
    \Theta(L_{\rm kink} - \delta),
\label{fluctuation_kink}
\end{equation}
where $L_{\rm kink}$ is a distance between kinks. The step
function $\Theta(L_{\rm kink} - \delta)$ represents
the effect that the fluctuation becomes negligible far
from the kink, and $\alpha_{\rm kink}$ represents
the kink configuration,
\begin{equation}
\alpha_{\rm kink}
= \hat{n} \cdot \vec{p}, ~~~
\vec{p}
= \left[ \frac{\vec{r}^{~\prime}}
              {|\vec{r}^{~\prime}|^2}
  \right]^{\sigma_{\rm kink}+0}_{\sigma_{\rm kink}-0},
\label{kink-amplitude}
\end{equation}
where $\sigma_{\rm kink}$ is the position of the kink and
$\vec{p}$ characterizes the change of the direction of the string
at the kink and represents the kink amplitude.

%%%%%%%%%%%%%%%%%%%%%%%%%%%%%%%%%%%%%%%%%%%%%%%%%%%%%%%%%%%%%%%%%
\section{Analytic model of cosmic string network \label{scaling}}
%%%%%%%%%%%%%%%%%%%%%%%%%%%%%%%%%%%%%%%%%%%%%%%%%%%%%%%%%%%%%%%%%

In this section, we develop an analytic model which describes
the behavior of a cosmic string network. First, the average
string length $\xi$ and the rms velocity $v_{\rm rms}$ are
calculated using a velocity-dependent one-scale model
\cite{Martins,Avgoustidis06}. Then, the number of kinks
in a horizon volume is calculated with an approach similar to
\cite{Allen} (see also \cite{Kibble91,Austin93,Vincent97}).

In the velocity-dependent one-scale model, a string network
is assumed to consist of straight string segments with the
average length $\xi$ and the rms velocity $v_{\rm rms}$.
This scale $\xi$ is also assumed to characterize the interstring
distance, that is, $\rho_{\rm seg} = \mu/\xi^2$, where
$\rho_{\rm seg}$ is the energy density of string segments,
respectively. In terms of $\xi$, the number of segments in a
horizon volume is expressed as
\begin{equation}
N_{\rm seg} = \frac{1}{\xi^3 H^3} = \gamma^3,
\end{equation}
where $H$ is the Hubble parameter and we defined
$\gamma \equiv 1/(\xi H)$.

The evolution of the network of segments is determined by
the cosmic expansion and the energy loss due to loop formation.
A loop formation can occur through the intercommutation of two
segments or the self-intercommutation of a single segment.
The characteristic timescale for loop formation is
$\sim \xi / P v_{\rm rms}$. For a universe with the scale
factor $a(t) \propto t^{\beta}$, the evolution equations
for $\gamma$ and $v_{\rm rms}$ are given by
\cite{Martins,Avgoustidis06}
\begin{eqnarray}
&& \frac{t}{\gamma} \frac{d\gamma}{dt}
   = 1 - \beta
     - \frac{1}{2} \beta \tilde{c} P v_{\rm rms} \gamma
     - \beta v_{\rm rms}^2,
\label{gamma_evolution} \\
&& \frac{dv_{\rm rms}}{dt}
   = (1-v_{\rm rms}^2) H
     \left[ k(v_{\rm rms}) \gamma - 2 v_{\rm rms} \right],
\label{v_evolution}
\end{eqnarray}
where $\tilde{c}$ is a constant which represents
the efficiency of the loop formation and
$k(v_{\rm rms}) \approx (2 \sqrt{2}/\pi)
(1 - 8 v_{\rm rms}^6)/(1 + 8 v_{\rm rms}^6)$
is the momentum parameter \cite{Martins}. Hereafter we
assume a matter-dominated universe and set $\beta = 2/3$.

It is known that a string network approaches a ``scaling"
regime where the characteristic scale grows with 
the horizon size~\cite{Kibble85}. This means that $\gamma$
and $v_{\rm rms}$ are asymptotically constant in time.
Here we assume that the scaling behavior is already
realized by the recombination time.
From (\ref{gamma_evolution}) and (\ref{v_evolution}),
we obtain the scaling values of $\gamma$ and $v_{\rm rms}$
neglecting their time derivatives. For small $\tilde{c}P$
they can be approximately given as,
\begin{equation}
v_{\rm rms}^2
\approx \frac{1}{2}
        - \frac{1}{2} \sqrt{\frac{\pi \tilde{c} P}{3 \sqrt{2}}},
~~~
\gamma = \frac{2 v_{\rm rms}}{k(v_{\rm rms})}
\approx \sqrt{\frac{\pi \sqrt{2}}{3 \tilde{c} P}}.
\label{solution}
\end{equation}
We see that small $P$, which means the inefficient loop
formation, leads to large $\gamma$ and hence large
$N_{\rm seg}$. From Eq. (\ref{solution}), we have the dependence
$\rho_{\rm seg} \propto P^{-1}$, which is consistent with
the result of numerical simulations in \cite{Sakellariadou}
while \cite{Avgoustidis06} obtained a relatively weaker
dependence on $P$. Actually there is no consensus on
the dependence on $P$ and we argue the effects of this
ambiguity on the pdf later.

Next, we consider the kink number evolution. Our approach
is based on the idea of \cite{Allen} although the formulation
is somewhat different. Small scale structure on strings
including kinks is also considered in \cite{Rocha} in
a different approach. Kinks are formed on string segments
when they intercommute and, simultaneously, some of
the existing kinks are removed through loop formation.
Furthermore, kinks decay due to stretching by the cosmic
expansion and the emission of gravitational waves.
Here we neglect the decay due to the gravitational wave
emission and focus on the decay due to cosmic expansion
since it is the most efficient decay process at
the matter-dominated stage~\cite{Allen}.

According to \cite{Bennett}, the kink amplitude, $p = |\vec{p}|$
(see Eq. (\ref{kink-amplitude})), decays with cosmic expansion
as $p(t) = p_{\rm f} (t/t_{\rm f})^{-\epsilon}$,
where $t_{\rm f}$ and $p_{\rm f}$ are the formation time and
the amplitude at the formation, respectively, and
\begin{equation}
\epsilon \equiv \frac{2(1-2v_{\rm rms}^2)}{3}
\approx \frac{2}{3}
        \sqrt{\frac{\pi \tilde{c} P}{3 \sqrt{2}}}\,.
\end{equation}
We count the number of kinks with amplitude
$p_{\rm min} \leq p \leq p_{\rm max}$ where $p_{\rm min}$
and $p_{\rm max}$ are free parameters. Later we show that
we need only the ratio $p_{\rm max}/p_{\rm min}$ for
our calculation and it will be determined by comparing our
pdf for $P=1$ with that obtained by the numerical simulations
\cite{Fraisse07}. Even if a kink is formed with
$p_{\rm f} > p_{\rm min}$, the cosmic expansion reduces
the amplitude gradually and  eventually it is no longer
counted as a kink after a time determined by
$p(t) = p_{\rm min}$. Therefore, the kink number in
a comoving volume $V(t) \propto a^3(t)$ is given
by the following integral of the formation rate
$d \bar{N}_{\rm form}(t,p)/dt dp$,
\begin{equation}
\bar{N}_{\rm kink}
= \int_{p_{\rm min}}^{p_{\rm max}} dp ~ \int_{t_0(p)}^{t} dt ~
  \frac{d \bar{N}_{\rm form}(t,p)}{dt dp},
\label{bar_N}
\end{equation}
where $t_0(p) = t (p/p_{\rm max})^{1/\epsilon}$, and a barred
quantity is a value in the comoving volume $V(t)$.

The formation rate of kinks, which is assumed here to be
independent of $p$, is proportional to the loop formation
rate, $d\bar{N}_{\rm loop}/dt$. Because the loop formation
rate determines the rate of the loss of the energy of
string segments, we have \cite{Allen}
\begin{equation}
\frac{d \bar{N}_{\rm form}(t)}{dt dp}
= \frac{q}{p_{\rm max}} \frac{d \bar{N}_{\rm loop}(t)}{dt}
= \frac{q \tilde{c} P v_{\rm rms}}{p_{\rm max} \alpha \xi}
  \frac{V}{\xi^3},
\end{equation}
where $q$ is a constant which represents the efficiency
of the kink formation and $\alpha$ is the average loop length
in units of $\xi$. Performing the integrations in (\ref{bar_N}),
the kink number in a horizon volume
$N_{\rm kink} = \bar{N}_{\rm kink}/(V H^3)$ is
\begin{equation}
N_{\rm kink}
\approx \frac{2 q \tilde{c} P v_{\rm rms} \gamma^4 \epsilon}
             {3 \alpha}
        \left( \frac{p_{\rm max}}{p_{\rm min}}
        \right)^{1/\epsilon},
\label{N_kink}
\end{equation}
where we have assumed $\epsilon \ll 1$. Because $N_{\rm kink}$
is independent of time, the kink number is also scaling.
This means that the average distance between kinks,
$L_{\rm kink}$, evolves in proportion to the horizon scale.
In fact $L_{\rm kink}$ is given by
\begin{eqnarray}
&& L_{\rm kink}
   \equiv \frac{N_{\rm seg} \xi}{N_{\rm kink}}
   = \frac{1}{K H}, \\
&& K \equiv \frac{N_{\rm kink} \gamma}{N_{\rm seg}}
     = \frac{N_{\rm kink}}{\gamma^2},
\label{K}
\end{eqnarray}
where the normalized linear kink density, $K$, is constant
in time. Thus we have expressed the numbers of string segments
and kinks in a horizon volume as functions of $P$.
%More realistic analysis
%with kink loss due to the loop formation and gravitational wave
%emission will be presented elsewhere \cite{Takahashi},
%although they are shown to be less important than cosmological
%expansion in the matter-dominant era \cite{Allen}.

%%%%%%%%%%%%%%%%%%%%%%%%%%%%%%%%%%%%%%%%%%%%%%%%%%%%%%%%%%%%%%%%%
\section{PDF of CMB fluctuations \label{pdf_CMB}}
%%%%%%%%%%%%%%%%%%%%%%%%%%%%%%%%%%%%%%%%%%%%%%%%%%%%%%%%%%%%%%%%%

Based on the elementary processes presented in section
\ref{fluctuation} and the network evolution model in section
\ref{scaling}, we calculate the pdf of the CMB temperature
fluctuations due to string segments and kinks.

A photon ray is scattered by segments many times through
its way from the last scattering surface to an observer.
Hence the temperature fluctuation would behave like
a random walk and the pdf from segments would be approximated
by the Gaussian distribution. If we treat a segment as
a particle with the cross section $\xi^2$ as we discussed
below Eq. (\ref{fluctuation_segment}), the optical depth is
\begin{equation}
\tau
= \int^{z_{\rm rec}}_{0} N_{\rm seg} H^3
  \xi^2 \frac{dz}{H (1+z)}
= \gamma \log{(1+z_{\rm rec})},
\end{equation}
where $z_{\rm rec} \approx 1100$ is the redshift
at the recombination. This is estimated as
$\gamma \log{(1+z_{\rm rec})} \approx 16$ for $P = 1$
and larger for smaller $P$. Although the temperature
change at each scattering is different depending on the
factors $\alpha_{\rm seg}$ and $v$, it would be a
good approximation to estimate the dispersion of the pdf
using their statistical averages. Therefore, remembering
Eq. (\ref{fluctuation_segment}), the dispersion is
evaluated as,
\begin{eqnarray}
\sigma
&=& \Delta_{\rm seg} \frac{\sqrt{\tau}}{2} \nonumber \\
&=& 2 \pi \frac{v}{\sqrt{1-v^2}} \alpha_{\rm seg} G \mu
    \sqrt{\gamma \log{(1+z_{\rm rec})}}
\nonumber \\
&\approx& 2 \pi \alpha_{\rm seg} \sqrt{\log{(1+z_{\rm rec})}}
          \left( \frac{\pi \sqrt{2}}{3 \tilde{c} P}
          \right)^{1/4}
          G \mu,
\label{sigma}
\end{eqnarray}
where we have set $v = v_{\rm rms}$ and substituted
Eq. (\ref{solution}) in the third equality, and
$\Delta_{\rm seg}$ and $\alpha_{\rm seg}$ should be understood
as their statistical averages, $\sqrt{<\Delta_{\rm seg}^2>}$
and $\sqrt{<\alpha_{\rm seg}^2>}$, respectively.
Here it should be noted that the PDF from string segments
should be, strictly speaking, the binomial distribution.
However, for the number of trials evaluated above ($\sim 16$),
the deviation of the binomial distribution from the Gaussian
distribution is negligibly small.

Next, let us consider the contribution from kinks.
The temperature fluctuation depends on the impact parameter
as given by (\ref{fluctuation_kink}). Solving for $\delta$
as a function of $\Delta$, we have,
\begin{eqnarray}
&& \delta(\Delta) = L_{\rm kink} e^{-|\Delta|/2\Delta_0}, \\
&& \Delta_0 \equiv 2 \alpha_{\rm kink} G \mu \label{Delta_0}.
\end{eqnarray}
Therefore the differential cross section with the temperature
fluctuation $\Delta$ can be written as
\begin{equation}
\frac{d \sigma_{\rm kink}}{d \Delta}
= \left| \frac{d}{d \Delta} \delta^2(\Delta) \right|
= \frac{L_{\rm kink}^2}{\Delta_0} e^{-|\Delta|/\Delta_0},
\end{equation}
where $\alpha_{\rm kink}$ should again be understood as
its statistical average of the kink configuration and
$\sigma_{\rm kink}$ should not be confused with
the coordinate on a string in section \ref{fluctuation}.
Then the pdf of temperature fluctuations due to kinks is
\begin{eqnarray}
\frac{d P_{\rm kink}}{d \Delta}
&=& \int_0^{z_{\rm rec}} N_{\rm kink} H^3
    \frac{d \sigma_{\rm kink}}{d \Delta}
    \frac{dz}{H(1+z)}
\nonumber \\
&=& \frac{\gamma^2}{K \Delta_0}
    e^{- |\Delta|/\Delta_0} \log{(1+z_{\rm rec})}.
\label{P_kink}
\end{eqnarray}
The normalization factor can be evaluated as
\begin{eqnarray}
A &\equiv& \frac{\gamma^2 \log{(1+z_{\rm rec})}}{K \Delta_{0}}
%\nonumber \\
%  &=& \frac{3 \alpha \log{(1+z_{\rm rec})}}
%           {2 q \tilde{c} P v \epsilon \Delta_{0}}
%      \left( \frac{p_{\rm max}}{p_{\rm min}}
%      \right)^{-1/\epsilon}
\nonumber \\
  &\approx& \frac{3 \pi \alpha \log{(1+z_{\rm rec})}}
                 {8 q \alpha_{\rm kink} G \mu}
            \left( \frac{3 \sqrt{2}}{\pi \tilde{c} P}
            \right)^{3/2}
            \left( \frac{p_{\rm max}}{p_{\rm min}}
            \right)^{-(3/2)\sqrt{3 \sqrt{2} / \pi \tilde{c} P}},
%\nonumber \\
%&& \approx 5.9
%           \alpha_{\rm kink}^{-1} P^{-3/2}
%           \left( \frac{p_{\rm max}}{p_{\rm min}}
%           \right)^{-2.4/\sqrt{P}}
%           (G \mu)^{-1},
\label{A}
\end{eqnarray}
where we have used Eqs. (\ref{solution}), (\ref{N_kink})
and (\ref{K}) in the second equality. Thus we have a pdf
of the form,
\begin{eqnarray}
&& \frac{dP_{\rm tot}}{d\Delta}
   = \frac{dP_{\rm G}}{d\Delta}
     + \frac{dP_{\rm NG}}{d\Delta},
\label{Ptot} \\
&& \frac{dP_{\rm G}}{d\Delta}
   = \frac{1}{\sqrt{2 \pi} \sigma}
     e^{-\Delta^2/2 \sigma^2},
\label{PG} \\
&& \frac{dP_{\rm NG}}{d\Delta}
   = A e^{-|\Delta|/\Delta_0},
\label{PNG}
\end{eqnarray}
where $\sigma$, $\Delta_0$ and $A$ are given by
Eqs. (\ref{sigma}), (\ref{Delta_0}) and (\ref{A}), respectively.
$dP_{\rm G}/d\Delta$ is the Gaussian part due to
frequent scatterings by string segments, and
$dP_{\rm NG}/d\Delta$ is the non-Gaussian tail
due to rare scatterings by kinks. Here, because
$dP_{\rm NG}/d\Delta \ll 1$ as we see just below,
we have normalized $dP_{\rm G}/d\Delta$ as
$\int_{-\infty}^{\infty} d\Delta ~ dP_{\rm G}/d\Delta = 1$.

In the limit $P \rightarrow 1$, we have
\begin{eqnarray}
&& \sigma \approx 14 G \mu, ~~~
   A \approx 10 \alpha_{\rm kink}^{-1}
             \left( \frac{p_{\rm max}}{p_{\rm min}}
             \right)^{-5.1}
             (G\mu)^{-1}, \nonumber \\
&& \Delta_0 = 2 \alpha_{\rm kink} G \mu,
\label{analytic}
\end{eqnarray}
where we have set $q = 2$, $\tilde{c} = 0.23$ and $\alpha = 0.1$
as their standard values \cite{simulation}. Here we used
$\alpha_{\rm seg} = 1/\sqrt{2}$ for the statistical average
assuming a random distribution of $\vec{r}^{~\prime}$ and
$\dot{\vec{r}}$. Contrastingly, the statistical average
of $\alpha_{\rm kink}$ cannot easily be obtained because
kinks evolve in time and their distribution is nontrivial.
This problem is closely related to the number count of kinks
discussed in section \ref{scaling} and we will postpone it
to the future work.

On the other hand, the pdf from numerical simulations
\cite{Fraisse07} can be also described as Eqs. (\ref{Ptot}),
(\ref{PG}) and (\ref{PNG}) with
\begin{equation}
\sigma_{\rm sim} \approx 12 G \mu, ~
A_{\rm sim} \approx 0.03 (G \mu)^{-1}, ~
\Delta_{0, {\rm sim}} \approx 9 G \mu.
\label{simulation}
\end{equation}
First, we note that the dispersion of the Gaussian part
is well reproduced without any adjustable parameters.
This would imply that our interpretation of the Gaussian
part as frequent scatterings by segments is reasonable.
Next, to compare the non-Gaussian part of Eqs.
(\ref{analytic}) with (\ref{simulation}), we must
specify the values of $\alpha_{\rm kink}$ and
$p_{\rm max}/p_{\rm min}$. To estimate these parameters
from the first principles is, however, beyond the scope of
the present paper. This will be discussed in a forthcoming
paper~\cite{Takahashi}. Here we just treat them as
phenomenological parameters and put
$\alpha_{\rm kink} = 4.5$ and $p_{\rm max}/p_{\rm min} = 2.3$
to make (\ref{analytic}) and (\ref{simulation}) consistent.

Before we discuss the $P$ dependence of the pdf,
let us consider the effect of finite angular resolution of 
CMB observation. As we saw above, a ray has to pass nearby
a kink to have a large temperature fluctuation but it may not
be resolved if the angular resolution is finite. This effect
can be taken into account by assuming that the impact
parameter $\delta$ cannot be smaller than a certain value
$\delta_{\rm min}(z,\theta)$ determined by the redshift of
a kink and the angular resolution $\theta$,
\begin{equation}
\delta_{\rm min}(z,\theta)
= \theta d_{\rm A}(z)
= 2 \theta H^{-1}(z) ( \sqrt{1+z} - 1),
\end{equation}
where $d_{\rm A}(z)$ is the angular diameter distance.
The largest fluctuation which can be generated by a kink at $z$
is, then,
\begin{eqnarray}
\Delta_{\rm max}(z,\theta)
&=& 2 \Delta_0
    \log{\left( \frac{L_{\rm kink}(z)}{\delta_{\rm min}(z,\theta)}
         \right)}
\nonumber \\
&=& - 2 \Delta_0
      \log{\left[ 2 \theta K \left( \sqrt{1+z} - 1 \right) \right]}.
\end{eqnarray}
Solving this in terms of $z$, we obtain the largest redshift of
kinks which contributes to a specific value of $\Delta$,
\begin{equation}
z_{\rm max}(\Delta,\theta)
= {\rm min} \left[
  \left( 1 + \frac{e^{-|\Delta|/2 \Delta_0}}{2 \theta K} \right)^2
  - 1,
  z_{\rm rec} \right],
\end{equation}
where this is bounded by $z_{\rm rec}$ because we are considering
the ISW effect. Then the pdf modified by a finite angular
resolution is obtained by changing the integration range of
Eq. (\ref{P_kink}) from $[0,z_{\rm rec}]$ to
$[0,z_{\rm max}(\Delta,\theta)]$,
\begin{eqnarray}
\frac{d P_{\rm kink}^{\rm res}}{d \Delta}
&=& \int_0^{z_{\rm max}(\Delta,\theta)} N_{\rm kink} H^3
     \frac{d \sigma_{\rm kink}}{d \Delta}
     \frac{dz}{H(1+z)}
\nonumber \\
&=& \frac{2 \gamma^2}{K \Delta_0} e^{-|\Delta|/\Delta_0}
\nonumber \\
&\times&
    \left[
      \log{\left( 1 + \frac{e^{-|\Delta|/2 \Delta_0}}{2 \theta K}
           \right)}
      + \frac{1}{2}
        \frac{e^{-|\Delta|/2 \Delta_0}}
             {e^{-|\Delta|/2 \Delta_0} + 2 K \theta}
    \right],
\nonumber \\
\end{eqnarray}
where we have assumed $z_{\rm max}(\Delta,\theta) < z_{\rm rec}$.
In the limit of an infinite resolution, this reduces to
Eq. (\ref{P_kink}), and in the opposite limit, we have
\begin{equation}
\frac{d P_{\rm kink}}{d \Delta}
\approx \frac{3 \gamma^2}{2 \theta K^2 \Delta_0}
        e^{-3 |\Delta|/2 \Delta_0}.
\end{equation}
The effect of a finite resolution is important for
$|\Delta| > \Delta_1$ where $\Delta_1$ is defined by
$e^{-\Delta_1/2 \Delta_0}/(2 \theta K) = 1$, that is,
$\Delta_1 = - 2 \Delta_0 \log{(2 \theta K)}$. Setting
$\theta = 0.42'$ which was adopted in \cite{Fraisse07},
$\Delta_1$ is estimated as $\approx 68 G \mu$
for $P = 1$ and larger for smaller $P$. Thus the non-Gaussian
tail steepens slightly for large $|\Delta|$ but the pdf
of \cite{Fraisse07} is still well reproduced.

\begin{figure}[t]
\includegraphics[width=85mm]{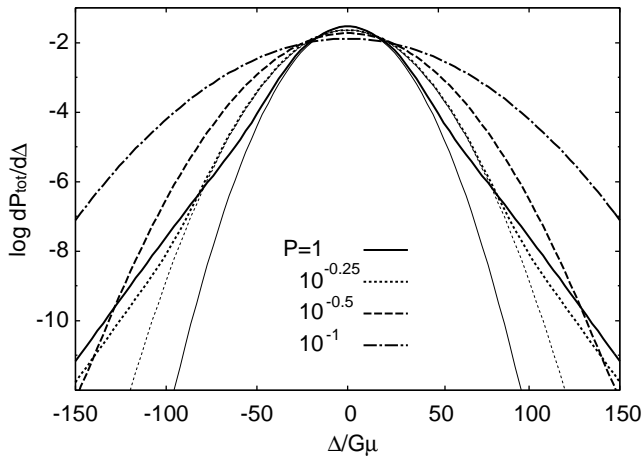}
\caption{Dependence of the pdf with angular resolution
$\theta = 0.42'$ on the intercommuting probability $P$
(thick lines). The respective Gaussian parts are plotted
with thin lines for comparison. For $P = 1$ and $10^{-0.25}$,
the pdfs deviate significantly from the Gaussian distribution.
For $P \lesssim 10^{-0.5}$, pdfs are almost Gaussian.}
\label{P-dependence}
\end{figure}

\begin{figure}[t]
\includegraphics[width=85mm]{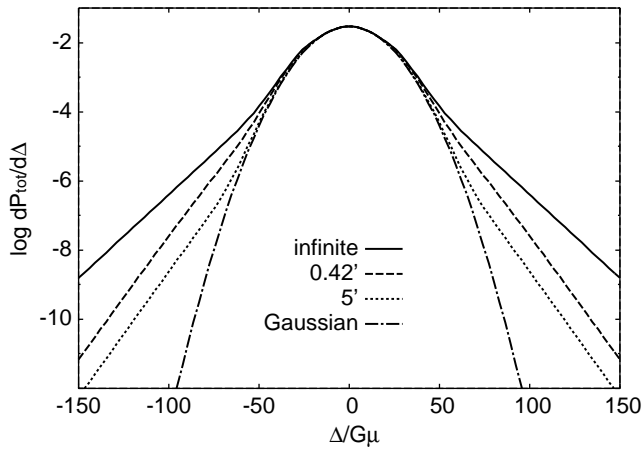}
\caption{Dependence of the pdf with $P = 1$ on angular
resolution. The Gaussian part is also plotted for
comparison.}
\label{resolution}
\end{figure}

In Fig. \ref{P-dependence}, the dependence of the pdf
on the intercommuting probability $P$ are shown
with angular resolution $\theta = 0.42'$.
The respective Gaussian parts are also plotted.
As we see, as $P$ decreases, the Gaussian dispersion
increases and the contribution of the non-Gaussian
tail is suppressed. For the case of $P=1$, the pdf
is almost Gaussian for $|\Delta| \lesssim 50 G \mu$
while non-Gaussian tails can be seen for larger $|\Delta|$.
Contrastingly, the pdfs are almost Gaussian for a wide
range of $|\Delta|$ in the cases with $P \lesssim 10^{-0.5}$.
Thus the non-Gaussianity could be a probe of the cosmic
string property, $P$.

Fig. \ref{resolution} shows the dependence on angular
resolution with $P=1$. We see that for a typical angular
resolution, $5'$, of future observations such as
{\it Planck} \cite{Planck}, non-Gaussian feature is
highly suppressed even for $P=1$ because kinks can
not be resolved. Thus we would need observations with
an arcminute resolution.

Note that our pdf is symmetric for positive and negative
$\Delta$ and cannot reproduce the non-zero skewness reported
in \cite{Fraisse07}. This is because we have assumed
the long segments to be straight. In fact, for a straight
segment, temperature fluctuation is symmetric between
positive and negetive values as is seen from
Eqs. (\ref{fluctuation_segment_atan}) and (\ref{alpha_seg}).
In this case, even if the number of scatterings is
relatively small and the Gaussian approximation is not
valid, skewness does not appear.

However, skewness would appear if we take the curvature
of segments and its correlation with velocity into account
\cite{Takahashi,Hindmarsh}. The reason is as follows.

First, if a segment has a curvature, the symmetry of
temperature fluctuation between positive and negative
values is broken. The asymmetry depends on the angle
between the velocity and curvature vectors. Therefore
the curvature cannot induce a skewness itself because
the asymmetry would be canceled out by multiple
scatterings of strings with various configuration
of velocity and curvature vectors. However, if there
is a correlation between velocity and curvature,
there would be a nonzero expectation value for
the asymmetry. This is the mechanism we believe
the skewness is induced from.

To estimate the skewness, we need to extend the
fluctuation formula Eq. (\ref{fluctuation_segment_atan})
taking the curvature of a segment into account.
With the nonzero expectation value of asymmetry,
the deviation from the Gaussian distribution would
lead to a nonzero skewness. The extention of 
Eq. (\ref{fluctuation_segment_atan}) and the
evaluation of the deviation from the Gaussian
distribution due to the finite number of scatterings
will be discussed in a separate article \cite{Takahashi}.
Nevertheless, it would be surprising that most of
the features of the pdf obtained by the numerical
simulations \cite{Fraisse07} can be interpreted
by our simple model with just straight segments and kinks.

%%%%%%%%%%%%%%%%%%%%%%%%%%%%%%%%%%%%%%%%%%%%%%%%%%%%%%%%%%%%%%%%%
\section{Discussion and Summary \label{summary}}
%%%%%%%%%%%%%%%%%%%%%%%%%%%%%%%%%%%%%%%%%%%%%%%%%%%%%%%%%%%%%%%%%

In this paper, we have computed analytically the pdf of
small-scale CMB temperature fluctuations due to cosmic
(super-)strings with a simple model with straight segments
and kinks. Our purposes were to interpret the results
of numerical simulations in \cite{Fraisse07} and
study the effect of the string intercommuting probability $P$.
We have combined and extended a velocity-dependent
one-scale model and a kink model to calculate the numbers of
string segments and kinks in a horizon volume consistently.
Thus obtained pdf consists of a Gaussian component due to frequent
scatterings by string segments and a non-Gaussian tail due to
close encounters with kinks. The dispersion of the Gaussian part
obtained by numerical simulations \cite{Fraisse07} is well
reproduced without any adjustable parameters. On the other hand,
the non-Gaussian tail contains two phenomenological parameters
and we determined them by comparing it with that of the numerical
result for $P=1$ by Fraisse et al. \cite{Fraisse07}. Then we
clarified the $P$ dependence of the pdf and found that
the non-Gaussian tail diminishes as $P$ decreases.

Let us argue the ambiguity in our string network model
in section \ref{scaling}. The evolution of large- and
small-scale structure has not been well understood either
analytically or numerically. In particular, the dependence
of $\gamma$, $v_{\rm rms}$ and $K$ on $P$ is quite important
to derive the $P$ dependence of the pdf by our formalism.
As we pointed out below Eq. (\ref{solution}), the dependence
of $\gamma$ in our network model is consistent with
that of \cite{Sakellariadou} while \cite{Avgoustidis06}
claims a relatively weaker dependence. However, they are
consistent in that a small $P$ results in large $\gamma$,
$N_{\rm seg}$ and $\rho_{\rm seg}$. Then, from the
second equation of Eq. (\ref{sigma}), it would be robust
that the dispersion of the Gaussian part increases as
$P$ decreases suppressing the non-Gaussian tail, assuming
that $v_{\rm rms}$ would not differ significantly from
$1/\sqrt{2}$ (see Eq. (\ref{solution})). Anyway,
the parameters of the pdf can be calculated by our
formalism once $\gamma$, $v_{\rm rms}$ and $K$ are given
as functions of $P$. Thus it would be interesting
to calculate the pdf using those functions obtained
by other models and numerical simulations.

Our pdf is contributed only from the ISW effect of cosmic strings.
Although the primary temperature fluctuations are substantially
damped at small scales we consider here ($\sim O(1)~{\rm min}$),
other secondary fluctuations such the as Sunyaev-Zel'dovich effect
would become important depending on the value of $G\mu$.
To discuss further observational prospects of cosmic
(super-)strings, it would be necessary to compare contributions
from various secondary fluctuations.

Also it is important to compute other observational quantities
with our simple formalism. In particular, the power spectrum
contributed from cosmic strings has been calculated by many
authors \cite{Hindmarsh94,Fraisse07,Pogosian08} and
the comparison with them would further allow us
to check the applicability of our formalism. As to
non-Gaussianity, observationally more interesting quantities
than the pdf would be higher-order correlation functions.
The work along this direction is in progress~\cite{Takahashi}.

%%%%%%%%%%%%%%%%%%%%%%%%%%%%%%%%%%%%%%%%%%%%%%%%%%%%%%%%%%%%%%%%%
\acknowledgements
%%%%%%%%%%%%%%%%%%%%%%%%%%%%%%%%%%%%%%%%%%%%%%%%%%%%%%%%%%%%%%%%%

This work is supported in part by Monbukagaku-sho Grant-in-Aid
for the global COE programs, "The Next Generation of Physics,
Spun from Universality and Emergence" at Kyoto University and
gQuest for Fundamental Principles in the Universe:
from Particles to the Solar System and the Cosmosh
at Nagoya University. YS and DY are supported by Grant-in-Aid
for JSPS Fellows. MS is supported by JSPS Grant-in-Aid for
Scientific Research (B) No.~17340075, and (A) No.~18204024,
and by JSPS Grant-in-Aid for Creative Scientific Research
No.~19GS0219.

\end{document}